\documentclass[final]{elsarticle}

\usepackage{inputenc} 
\usepackage{graphicx}
\usepackage[top=20mm,bottom=20mm,left=20mm,right=20mm]{geometry}
\usepackage{amsmath}
\usepackage{mathtools}
\usepackage{xcolor}
\usepackage{ragged2e}
\usepackage{subcaption}
\usepackage{float}
\usepackage[breaklinks]{hyperref}
\usepackage{amssymb}

\let\oldref\ref
\renewcommand{\ref}[1]{(\oldref{#1})}

\newtheorem{defn}{Definition}[section]
\newtheorem{rem}{Remark}[section]

\journal{Journal}

\begin{document}

\begin{frontmatter}

\title{Basic and extendable framework for effective charge transport in electrochemical systems}

\author[label1]{Jeta Molla\corref{cor1}}
\ead{jm188@hw.ac.uk}

\author[label1]{Markus Schmuck}
\ead{schmuck@compsyst.com}
\ead[url]{compsyst.com}
\address[label1]{Maxwell Institute for Mathematical Sciences and\\ 
School of Mathematical and Computer Sciences,\\ 
Heriot-Watt University,\\ 
EH144AS, Edinburgh, UK}

\cortext[cor1]{Corresponding author}

\begin{abstract}
We consider basic and easily extendible transport formulations for lithium batteries consisting of an anode (Li-foil), a separator (polymer electrolyte), 
and a composite cathode (composed of electrolyte and intercalation particles). Our mathematical investigations show the following novel features: 
(i) \emph{complete and very basic description of mixed transport processes} relying on a neutral, binary symmetric electrolyte resulting in a non-standard Poisson equation for the electric potential together with interstitial diffusion approximated by classical diffusion; (ii) \emph{ upscaled and basic composite cathode equations allowing to take geometric and material features of electrodes into account}; 
(iii) \emph{the derived effective macroscopic model can be numerically solved with well-known numerical strategies for homogeneous domains} and hence does not require to solve a high-dimensional numerical problem or to 
depend on a computationally involved multiscale discretisation strategies where highly heterogeneous and realistic, nonlinear, and reactive boundary conditions are still unexplored. We believe that the here proposed basic and easily 
extendible formulations will serve as a basic and simple setup towards a systematic theoretical and experimental understanding of complex electrochemical systems and their optimization, e.g. Li-batteries.
\end{abstract}

\begin{keyword}
lithium batteries, multiscale modeling, Butler-Volmer equations, homogenization, electrode design
\end{keyword}

\end{frontmatter}


\section{Introduction}

Energy storage systems play an increasingly important role for reliable, efficient, and preferably green energy and delivery in developed countries and also between them. 
Two major developments make affordable and endurable 
energy storage a necessity: (i) the global awareness of 
climate change and as such the need for renewable and low 
CO$_2$ energy-consumption/production; (ii) the realisation and affordability of electric mobility (cars and buses). 
In order to make storage systems more affordable, it is 
important to have a proper physical, chemical, and 
mathematical understanding of the processes involved in 
order to give systematic (i.e. based on 
variational principles) guidance on design optimization. 
Since electric cars are expected to become a multi-billion dollar buisness until 2030 and Li-ion batteries play a 
major role in this development, we aim here to provide a
fundamental, basic, and effective macroscopic description of an 
active electrode. 

Due this expected demand, recently an increasing interest in mathematical modeling of lithium batteries emerged. Well-known and commonly used macroscale models were developed by Newman and 
collaborators already decades ago, e.g. \cite{newman} and \cite{a14}, which serves as basis for the present investigations. In order to improve the battery performance, it will be crucial to connect material properties and the geometry of microstructure to current-voltage characteristics. This has recently led to an increased interest in the systematic derivation of effective macroscopic charge transport equations \cite{Schmuck2011,Richardson2012}. In fact, for the full nonlinear Poisson-Nernst-Planck equations first rigorous error estimates have been derived in \cite{Schmuck2012}. Related research for porous and heterogeneous media are \cite{Timofte2014,Allaire2017,Schmuck2019} for instance. 

\begin{figure} 
\centering
\includegraphics[width=17cm]{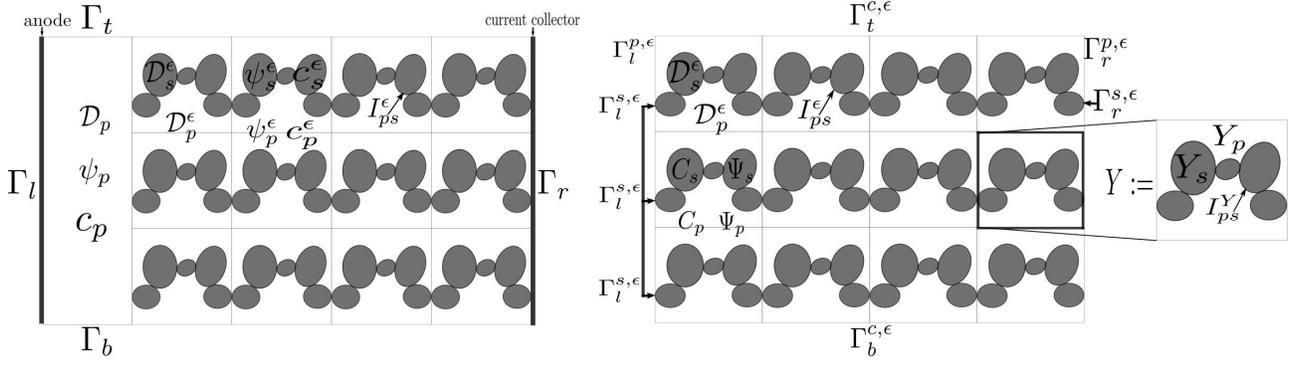}
\caption{{\bf Left:}  Schematic of lithium battery with separator ${\cal D}_p$, (homogenized) composite cathode ${\cal D}_c$, anode $\Gamma_l$,  and cathodic current collector $\Gamma_r$.   {\bf Right:} Microscopic composite cathode ${\cal D}_c:={\cal D}_p^\epsilon\cup{\cal D}_s^\epsilon$ as a periodic extension of a reference cell $Y:=Y_p\cup Y_s$ of length $\ell$. 
}
\label{fig:battery}
\end{figure}

We consider 
a basic and easily extendible non-re-chargeable Li-battery consisting of a polymer electrolyte/\linebreak separator ${\cal D}_p$, a composite cathode ${\cal D}_c$, a 
Li-foil $\Gamma_l$ as anode, 
and a cathodic current collector $\Gamma_r$, see Fig. \ref{fig:battery} (Left). The composite cathode ${\cal D}_c$ can be 
identified as the periodic extension of 
a statistically defined, characteristic reference cell $Y$ of length $\ell$, see Fig. \ref{fig:battery} (Right).
This leads to a so-called 
heterogeneity parameter 
$\epsilon:=\frac{\ell}{L}$ where $L$ is the length of the cathode.
Hence, ${\cal D}_c$ is highly heterogeneous and composed of an electrolyte ${\cal D}_p^\epsilon$ and a solid intercalation phase ${\cal D}_s^\epsilon$ such that 
${\cal D}_c:={\cal D}_p^\epsilon\cup{\cal D}_s^\epsilon$. The interface between the polymer and solid phase is denoted by $I_{ps}^\epsilon:=\partial {\cal D}_p^\epsilon\cap\partial {\cal D}_s^\epsilon$. We model Li-diffusion in 
neutral, binary symmetric electrolytes 
by the dilute solution theory \cite{newman} and 
Li-transport in solid intercalation hosts is described by classical diffusion. Based on this basic formulation, we systematically derive effective 
macroscopic cathode equations using the method of 
asymptotic two-scale expansions \cite{a11}. 
 The central quantities of interest are 
 the evolution of Li-density in 
 dependence of an applied electrical 
 current $I_a$ and the resulting electric potential. 
As shown in Fig.~\ref{fig:battery}, we apply the following notation 
for the Li-density $c$ and the 
potential $\psi$ in various domains 
$\mathbb{D}\in\{{\cal D}_p,{\cal D}_p^\epsilon,{\cal D}_s^\epsilon\}$, i.e.,
\begin{equation}
c :=
\begin{cases}
c_p\,,
    & \text{in } 
    \mathbb{D}={\cal D}_p
    \,,
\\
c^\epsilon_p\,,
    & \text{in }
    \mathbb{D}={\cal D}^\epsilon_p
    \,,
\\
c^\epsilon_s\,,
    & \text{in }
    \mathbb{D}={\cal D}^\epsilon_s
    \,.
\end{cases}
\qquad\qquad
\psi:=
\begin{cases}
\psi_p\,,
    & \text{in } 
    \mathbb{D}={\cal D}_p
    \,,
\\
\psi^\epsilon_p\,,
    & \text{in }
    \mathbb{D}={\cal D}^\epsilon_p
    \,,
\\
\psi^\epsilon_s\,,
    & \text{in }
    \mathbb{D}={\cal D}^\epsilon_s
    \,.
\end{cases}
\end{equation}
Under an applied discharging current density $i_a:=I_a/|\Gamma^{s,\epsilon}_r|$, charge transport in the homogeneous 
and heterogeneous electrolyte phases 
$\mathbb{D}={\cal D}_p$ 
and 
$\mathbb{D}={\cal D}_p^\epsilon$, 
respectively, is governed by
\begin{flalign} \label{polymer}
 \begin{cases}
 \frac{\partial c}{\partial t } = \Delta c \quad 
 & \text{in} \;\mathbb{D}\,,
\\\quad 
\nabla c \cdot { \bf n} =g    \quad & \text{on} \; 
\partial\mathbb{D}\,,
\\
-\text{div}\left( c\nabla \psi\right)
	=-{\cal R}\Delta c\quad & \text{in} \; \mathbb{D},
\\\quad 
\psi
    = h_D 
    & \text{on} \; \Gamma_D^p,
 \\\quad  
 \nabla \psi\cdot {\bf n}
    = 
    h_N
    \quad & \text{on} \; \Gamma_N^p=\partial\mathbb{D}\setminus\Gamma_D^p
    \,,
\end{cases} 
\end{flalign}
where 
${\bf n}$ is an outward pointing normal vector, $\varepsilon_p$ and $\varepsilon_s$ are the electrical permittivities of ${\cal D}_p^\epsilon$ and ${\cal D}_s^\epsilon$, respectively, ${\cal R}$ is $\frac{D_+-D_-}{(z_+M_+-z_-M_-)F}$ with $D_i$, $M_i$, and $z_i$ being diffusion, mobility, and charge number of species $i\in\{+,\,-\}$. The symmetry assumption on the electrolyte implies $z_+=-z_-$. 
The Li density $c_p$ and electric potential $\psi_p$ in $\mathbb{D}={\cal D}_p$ satisfy the same system (\ref{polymer}) as 
$c_p^\epsilon$ and $\psi_p^\epsilon$ 
in $\mathbb{D}={\cal D}_p^\epsilon$ for boundary conditions defined as follows 
\begin{equation}
g:= 
\begin{cases}
\beta_lR_{BV}^l
    & \text{on }\Gamma_l
    \,,
\\
0
    & \text{on }
    \Gamma_t\cup\Gamma_b
    \,,
\\
\nabla c_p^\epsilon\cdot{\bf n}
    & \text{on }
    \Gamma_l^{p,\epsilon}\cap \partial\mathbb{D}
    \,,
\\
\beta_lR_{BV}^{ps}
    & \text{on }I^\epsilon_{ps}
    \,,
\\
0
    & \text{on }
    \Gamma_t^{c,\epsilon}
    \cup
    \Gamma_r^{p,\epsilon}
    \cup
    \Gamma_b^{c,\epsilon}
    \,,
\end{cases}
\hspace{-0.15cm}
h_D:=
\begin{cases}
\psi_a-\eta_a
    & \text{on }\Gamma_l
    \,,
\end{cases}
\hspace{0.2cm}
h_N:=
\begin{cases}
\frac{\varepsilon_s}{\varepsilon_p} \nabla \psi_s^\epsilon\cdot {\bf n}
	& \text{on }I^\epsilon_{ps}\cup(\Gamma_l^{s,\epsilon}\cup\partial{\cal D}_p)
	\,,
\\
\nabla\psi_p\cdot {\bf n}
    & \text{on }\Gamma_l^{p,\epsilon}\cap\partial{\cal D}_p
    \,,
\\
0
	&\text{on }\Gamma_b\cup\Gamma_t
	\,,
\\
0
	&\text{on }\Gamma_b^{c,\epsilon}\cup\Gamma_t^{c,\epsilon}\cup\Gamma_r^{p,\epsilon}
	\,,
\end{cases}
\end{equation}
where 
\begin{flalign}
i^{ps}_{BV}=i_{ps}R^{ps}_{BV}=i_{ps}\left(e^{\frac{\alpha_a F}{R T}\eta_\epsilon}-(c_s^m-c_s^\epsilon)e^{-\frac{\alpha_c F}{R T}\eta_\epsilon}\right),
\label{iBVps}
\end{flalign}
describes Butler-Volmer (BV) reactions 
across the interface $I^\epsilon_{ps}$ 
and 
\begin{flalign}
i^l_{BV}=  i_{ l}R^{l}_{BV}=i_l\left( e^{ \frac{\alpha_a F}{RT}\eta_a}-e^{-\frac{\alpha_c F}{RT}\eta_a}\right)
	\,,
\label{iBVl}
\end{flalign}
electrochmical reactions at the anode-electrolyte interface 
$\Gamma_l$. 
The exchange current densities in the BV equations (\ref{iBVps})--(\ref{iBVl}) are
$i_{ps}=Fk_{ps} \left(c_p^m-c_p^\epsilon\right)^{\alpha_c} c_p^{\alpha_a}$ 
and 
$i_l=F k_{a}^{\alpha_a}k_c^{\alpha_c}\left( c_p^m-c_p^\epsilon\right)^{\alpha_a}(c_p^\epsilon)^{\alpha_c}$ 
where $ k_{a}^{\alpha_a}$ and $k_c^{\alpha_c}$ are anodic and cathodic reaction rates, respectively. 
The parameters $\alpha_a$ and $\alpha_c$ are anodic and cathodic transfer coefficients, respectively, and $c_p^m$ and $c_s^m$ are  the maximum lithium densities in ${\cal D}_p^\epsilon$ and ${\cal D}_s^\epsilon$, respectively. 
Moreover, $\eta_a := \psi_a - \psi_p $ is the local value of the surface overpotential and $\psi_a$ denotes 
the potential at the anode (here simply Li-foil) and similarly, for the the equilibrium potential $U$ 
the overpotential across $I^\epsilon_{ps}$ is 
$\eta^\epsilon:=\psi_s^\epsilon-\psi_p^\epsilon - U$. 
Furthermore, the parameters $\beta_p:=\frac{i_{ps}L_{\rm ref}}{c_{\rm ref} D_p}$ and  $\beta_l:=\frac{i_lL_{\rm ref}}{c_{\rm ref} D_p}$ make the Butler-Volmer equations dimensionless for a reference length $L_{\rm ref}$, a reference concentration $c_{\rm ref}$, and Li-diffusion constant $D_p$ in the electrolyte.

It remains to describe electron and Li transport in ${\cal D}_s^\epsilon$, i.e.,
\begin{flalign}
\begin{cases}
a_1\frac{\partial c_s^\epsilon}{\partial t}=\Delta c_s^\epsilon \quad \quad& \text{in} \; {\cal D}_s^\epsilon, \\
 \quad \nabla c_s^\epsilon \cdot {\bf n}=-\epsilon\beta_{s} R^{ps}_{BV} \quad \quad& \text{on} \; I_{ps}^\epsilon,\\
 \quad \nabla c_s^\epsilon \cdot {\bf n}=0 \quad\quad & \text{on}  \; \Gamma_r^{s,\epsilon}, \\
 -\text{div}\left( \sigma_s \nabla \psi_s^\epsilon\right) =0 \quad \quad& \text{in} \; {\cal D}_{s}^\epsilon,\\
 \quad \sigma_s \nabla \psi_s^\epsilon\cdot {\bf n}= \epsilon\beta_{\psi} R^{ps}_{BV}\quad\quad & \text{on} \; I_{ps}^\epsilon,\\
 \quad \sigma_s\nabla\psi_s^\epsilon\cdot {\bf n}= a_sI_a \quad \quad & {\rm on} \;\Gamma^{s,\epsilon}_r.\\
\end{cases}
\label{classicaldiffusion}
\end{flalign}
The Li-diffusion times $\tau_s:=\frac{L_{\rm ref}^2}{D_s}$ and $\tau_p:=\frac{L_{\rm ref}^2}{D_p}$ in the solid phase and the polymer/electrolyte phase, respectively, define the dimensionless parameter $a_1:=\tau_s/\tau_p$ 
for Li-diffusion constants $D_s$ and $D_p$ in ${\cal D}_s^\epsilon$ and $\mathbb{D}\in\left\{ {\cal D}_p,\, {\cal D}_p^\epsilon \right\}$, respectively.  
$\sigma_s$ is the electrical conductivity of ${\cal D}_s^\epsilon$. The parameters $\beta_s:=\frac{i_{ps}L_{\rm ref}}{c_{\rm ref} D_s}$ and $\beta_\psi:=\frac{i_{ps}L_{\rm ref}}{\sigma_{\rm ref}}\frac{F}{RT}$ 
with $a_s:=\frac{L_{\rm ref}}{\sigma_{\rm ref} |\Gamma_r^{s,\epsilon}|}\frac{F}{RT}$ 
make the (\ref{iBVps}) after upscaling dimensionless. Finally, $\sigma_{\rm ref}$ is a reference conductivity. 

Of central interest in battery modelling and optimization is the effective macroscopic description of electrodes. To this end, 
we provide a systematic upscaling framework for active electrodes such as ${\cal D}_c={\cal D}_p^\epsilon\cup{\cal D}_s^\epsilon$ by passing to the limit $\epsilon \to 0$ 
and by relying on crucial microscopic ingredients via (\ref{polymer})--(\ref{classicaldiffusion}) 
such as geometry and specific material characteristics. The homogenization is explained 
in Section \ref{derivation} and the results are stated in the next section. 

\section{Main results}
\label{section 2}
Our main results depend on the following  well-accepted concept of local equilibrium \cite{lte2, lte1}. 

\begin{defn} \label{lte}
The chemical potential $\mu_p(C_p,\Psi_p)=\log C_p-{\cal R}\Psi_p$ is said to be in \emph{local thermodynamic equilibrium}, if and only if it holds that
\begin{flalign} 
\frac{\partial \mu_p(C_p,\Psi_p)}{\partial x_k}=\begin{cases}
0 &\qquad {\text in} \;Y_p,\\
\quad\\
\frac{\partial \mu_p(C_p,\Psi_p)}{\partial x_k} & \qquad {\text in }\;{\cal D}_c, 
\end{cases} 
\end{flalign}
for every $k\in\mathbb{N}$, $1\leq k \leq d,$ and the upscaled quantities $\{ C_p,\Psi_p\} $, which are independent of the microscale ${\rm \bf y}\in Y.$ 
\end{defn}
\begin{rem}
 \rm Local thermodynamic equilibrium is used in many different applications \cite{a1, sb, a9}. Definition \ref{lte} accounts for the fact that the macroscopic variables are so slow compared to the fast processes on the microscale (fast scale ${\bf y}:={\bf x}/\epsilon\in Y_p$) that their variations disappear thereon 
but not so on the (slow) macroscale ${\bf x}\in {\cal D}_c$. 
\end{rem}
Note that after upscaling the phases ${\cal D}_p^\epsilon$ and ${\cal D}_s^\epsilon$ are super-imposed on the whole composite cathode ${\cal D}_c$ while preserving the corresponding volume fractions. The specific boundaries are defined in Fig. \ref{fig:battery}. 

\medskip

{\bf Main results:} (Upscaled cathode equations) \emph{Under local thermodynamic equilibrium (Definition \ref{lte}), the microscopic formulations (\ref{polymer}) and (\ref{classicaldiffusion})
turn after upscaling into the following effective composite cathode formulations
\begin{flalign} \label{main1}
\begin{cases}
p\frac{\partial C_p}{\partial t}={\rm div}\left( \hat{D}_p \nabla C_p \right)+ \bar{\beta}_pR^{ps}_{BV}&\quad  {\text in} \; {\cal D}_c,\\
\quad \hat{D}_p\nabla C_p \cdot {\bf n}=\nabla c_p \cdot {\bf n} & \quad  {\text on} \; \Gamma_l^c,\\
 \quad \hat{D}_p\nabla C_p \cdot {\bf n}=0 & \quad  {\text on} \; \Gamma^c\setminus\Gamma_l^c,\\
 -{\rm div}\left( C_p\hat{D}_{\psi_p}\nabla \Psi_p\right)={\cal R}{\rm div}\left( \hat{D}_p  \nabla C_p \right)   & \quad  {\text in} \; {\cal D}_c,\\
 \quad \hat{D}_{\psi_p}\nabla \Psi_p\cdot{\bf n}=\nabla \psi_p \cdot {\bf n} &\quad  {\text on} \;\Gamma_l^c,\\
\quad C_p\hat{D}_{\psi_p}\nabla \Psi_p\cdot{\bf n}=0 &\quad  {\text on} \;\Gamma^c\setminus\Gamma_l^c,\\
\end{cases} 
\end{flalign}
and 
\begin{flalign} \label{main2}
\begin{cases}
q\frac{\partial C_s}{\partial t}={\rm div}\left(\hat{D}_s \nabla C_s\right)+\bar{\beta}_sR^{ps}_{BV}&\quad  {\text in}\;{\cal D}_c,\\
 \quad \hat{D}_{s}\nabla C_s\cdot{\bf n}=0& \quad  {\text on} \; \Gamma^c,\\
 -{\rm div}\left(\hat{\Sigma}\nabla\Psi_s\right)=\bar{\beta}_\psi R_{BV}^{ps} & \quad{\text in}\;{\cal D}_c,\\
 \quad  \hat{\Sigma}\nabla \Psi_s\cdot {\bf n}=a_sI_a& \quad  {\text on} \; \Gamma_r^c,\\
 \quad \hat{\Sigma}\nabla \Psi_s\cdot {\bf n}= 0& \quad  {\text on} \;\Gamma^c\setminus\Gamma_r^c,
 \end{cases}
\end{flalign}
where 
$p=\frac{|Y_p|}{|Y|}$, $q=a_1(1-p)$, $\bar{\beta}_{p}=|\Lambda|\beta_p$, $\bar{\beta}_{s}=|\Lambda|\beta_s$,  $\bar{\beta}_{\psi}=|\Lambda|\beta_\psi$, and $|\Lambda|=\frac{|I_{ps}^Y|}{|Y|}$. The effective material tensors $\hat{D}_p=\{\bar{d}^p_{ik} \}_{i,k=1}^{d}$, 
 $\hat{D}_{\psi_p}=\{\bar{d}^{\psi_p}_{ik} \}_{i,k=1}^{d}$, $\hat{D}_s=\{\bar{d}^s_{ik} \}_{i,k=1}^{d}$, and $\hat{\Sigma}=\{\bar{s}_{ik} \}_{i,k=1}^{d}$    are given by
 \begin{flalign}
 \bar{d}_{ik}^w =\frac{1}{|Y|}\sum\limits_{j=1}^{d}\int_{Y_w}\left( \delta_{ik}-\delta_{ij}\frac{\partial\xi^k_{w}}{\partial y_j}\right)\,\mathrm{d \bf y},\quad \bar{s}_{ik} =\frac{1}{|Y|}\sum\limits_{j=1}^{d}\int_{Y_s}\sigma_s\left( \delta_{ik}-\delta_{ij}\frac{\partial\xi^k_{\psi_s}}{\partial y_j}\right)\,\mathrm{d \bf y},
 \end{flalign}
for $w\in\{p, \psi_p, s\}$, $Y_{\psi_p}=Y_{p}$, and $Y_{\psi_s}=Y_s$.  The correctors $\xi^k_{m}({\bf y})$, $m \in\{p, \psi_p, s, \psi_s\}$, $\;1\leq k\leq d$, solve the following reference cell problems
\begin{flalign}
\qquad \xi^k_{m}:
\begin{cases} 
-\sum\limits_{i,j=1}^{d}\frac{\partial}{\partial y_i}\left(\delta_{ij}\frac{\partial \xi^k_{m}}{\partial y_j} -{\bf e}_k\right)
= 0 
\quad  {\text in} \;  Y_m,\\
\qquad \big(\nabla \xi^k_{m}-{\bf e}_k \big)\cdot {\bf n}=0 \;{\text on}\; I_{ps}^Y\; 
\\
{\text and }\; \xi_{m}^k\text{ is }Y_m\text{-periodic with $\int_{Y} \xi_m^k\,d{\bf y}=0$}.
\end{cases}
\label{cellp}
\end{flalign} 
}

A more detailed discussion and extensions will appear in \cite{Schmuck2020book}.

\section{Derivation of effective macroscopic equations} \label{derivation}
The diffusion and elliptic equations, e.g. (\ref{polymer})$_1$--(\ref{polymer})$_2$ and (\ref{classicaldiffusion}) are standard in homogenization theory (see  \cite{a11,a15,a10,homogenization} for instance). However, equation (\ref{polymer})$_5$ shows an unexpected form due electro-neutrality and therefore we state the relevant steps of the 
derivation. With the asymptotic two-scale expansions \cite{a10,homogenization}
\begin{flalign} \label{asymptotic2}
 \qquad \begin{split}
 u^{\epsilon}(t,{\bf x})= u(t,{\bf x}, {\bf y})=  U(t,{\bf x},{\bf y}) + \epsilon u^1(t,{\bf x},{\bf y})+ \epsilon^2 u^2(t,{\bf x},{\bf y}) + \dots, \quad {\rm for}\; u\in\{c_p,\psi_p, \psi_s\}
 \end{split} &&
\end{flalign} 
and the following operators
\begin{flalign}
\qquad \begin{aligned}
&A_0= -{\cal R}L_{yy}(1) , \qquad \qquad  & B_0&= -L_{yy}(C_p),\\
&A_1= - {\cal R}\left[L_{xy}(1)+L_{yx}(1) \right],\qquad \qquad  &  B_1&= - \left[ L_{xy}(C_p) +L_{yx}(C_p)+L_{yy}(c_p^1) \right],  \\
&A_2= - {\cal R}L_{xx}(1), \qquad \qquad &  B_2&= - \left[ L_{xx}(C_p)+L_{xy}(c_p^1)+L_{yx}(c_p^1)+L_{yy}(c_p^2)\right],
\end{aligned}
\label{op2}
\end{flalign}
where $L_{xy}(u)=\sum_{i,j=1}^{d} \frac{\partial}{\partial x_i}\left(u \delta_{ij}\frac{\partial}{\partial y_j} \right)$, 
we obtain after collecting terms of equal power in $\epsilon$ the following problems
\begin{flalign} \label{p1}
\qquad \mathcal{O}(\epsilon^{-2}): \begin{cases} 
B_0\Psi_p=-A_0C_p\qquad &\text{in} \; Y_p,\\
 \quad \nabla_{\bf y} \Psi_p\cdot {\bf n}=0\;{\rm on}\; I^Y_{ps}\; {\rm and } \;\Psi_p\;{\rm is}\;Y_p-{\rm periodic},\\
\end{cases}
\end{flalign}

\begin{flalign} 
\qquad \mathcal{O}(\epsilon^{-1}):\begin{cases} 
B_0\psi^1_p+B_1\Psi_p= -A_0c_p^1-A_1C_p\qquad &\text{in} \; Y_p,\\
\quad \nabla_{\bf y} \psi_p^1\cdot {\bf n}=-\nabla_{\bf x} \Psi_p\cdot {\bf n}\;{\rm on}\; I^Y_{ps}\; {\rm and } \;\psi^1_p\;{\rm is}\;Y_p-{\rm periodic},\\
\end{cases}
\label{p2}
\end{flalign}

\begin{flalign}
\qquad \mathcal{O}(\epsilon^{0}): \begin{cases} 
B_0\psi^2_p+A_0c_p^2
	=
	-B_2 \Psi_p-B_1\psi_p^1-A_1c^1_p-A_2C_p\qquad &\text{in} \; Y_p,
\\\quad 
\nabla_{\bf y} \psi_p^2\cdot {\bf n}
	-\nabla_{\bf x} \psi_p^1\cdot {\bf n}
	=
	\frac{\varepsilon_s}{\varepsilon_p}\left(\nabla_{\bf y} \psi_s^1
	+\nabla_{\bf x}\Psi_s\right)\cdot {\bf n}
	\;{\rm on}\; I^Y_{ps}
\\\quad
	{\rm and } \;\psi^2_p\;{\rm is}\;Y_p-{\rm periodic}.
\end{cases}
\label{p3}
\end{flalign}
System (\ref{p1}) immediately implies independence of $\Psi_p$ on the microscale ${\bf y}\in Y_p$. This motivates to make the following ansatz
\begin{flalign} \label{psi1}
 \quad \psi_p^1=-\sum_{k=1}^{d}\xi_{\psi_p}^k({\bf y})\frac{\partial \Psi_p}{\partial x_k},
\end{flalign}
which after inserting into (\ref{p2}) together with Definition \ref{lte} leads to the following cell problem for $1\leq k\leq d$, i.e.,
\begin{flalign} \label{cellp2}
\qquad \xi^k_{\psi_p}:\begin{cases} -\sum\limits_{i,j=1}^{d}\frac{\partial}{\partial y_i}\left(\delta_{ij}\frac{\partial \xi^k_{\psi_p}}{\partial y_j} -{\bf e}_k\right)
= 0 \quad  {\rm in} \;  Y_p,
\end{cases}
\end{flalign}
with boundary conditions as stated in (\ref{cellp}). 
Finally, we derive the effective macroscopic equation for    $\Psi_p$ via the Fredholm alternative \cite{a5, a12}. That means, problem (\ref{p3}) has a unique solution if it holds that 
 \begin{flalign} \label{p6}
 \qquad \int_{Y_p} \left[-B_1\psi_p^1-B_2\Psi_p-A_1c_p^1-A_2C_p \right] \mathrm{d \bf y}
 	= 0,
 \end{flalign}
where we already neglect possible boundary contributions which will disappear after rewriting. 
Using (\ref{cellp2}), the well-known standard definition for $\hat{D}_p$, 
and after 
defining the tensor $\hat{D}_{\psi_p}=\{ d_{ik}^{\psi_p} \}_{i,k=1}^{d}$ by $d_{ik}^{\psi_p}= \sum\limits_{j=1}^{d}\int_{Y_p}\left( \delta_{ik}-\delta_{ij}\frac{\partial\xi^k_{\psi_p}}{\partial y_j}\right)\,\mathrm{d \bf y}$, 
 allows us to rewrite (\ref{p6}) as the following homogenized equation for the associated electrical potential $\Psi_p$, i.e., 
 \begin{flalign} 
\qquad-\sum_{i,k=1}^{d}\frac{\partial}{\partial x_i}\left( C_p\hat{D}_{\psi_p}\frac{\partial \Psi_p}{ \partial x_k} \right)={\cal R}\sum_{i,k=1}^{d}\frac{\partial}{\partial x_i}\left( \hat{D}_p\frac{\partial C_p}{ \partial x_k} \right).
 \end{flalign}
 
\section{Conclusions} \label{Conclusion} 
We have established a basic charge transport formulation capturing the essential electrochemical 
features of lithium batteries (i.e., non-rechargeable) with the goal of having a convenient and easily 
extendible prototype framework for the investigation of the influence of active 
electrode materials (here the composite cathode). We believe that the presented results (upscaled formulation) 
allow to study the influence of material and geometric properties on the current-voltage behaviour of 
Li-batteries and provide also the fundamental basis for subsequent extensions towards modelling 
of ageing and cycling dynamics. 
In fact, the formulation introduced will be of interest to researchers doing battery modelling as we provide a 
complete set of boundary conditions for a general prototype model allowing for various extensions such as 
an active anode, different reaction models as well as extensions for ageing and cycling dynamics. 
Finally, this novel model framework relies on basic physcial and electrochemical principles and 
hence serves as a promising theoretical and 
efficient computational tool to investigate Li-batteries. From a computational point of view, it allows us to apply 
powerful numerical strategies well-known and developed for 
homogeneous domains in contrast to a possible multiscale discretization strategy requiring 
demanding implementations for boundary conditions on interfaces due to highly heterogeneous domains which 
itself imply costly constraints such as small enough mesh sizes.

\medskip

{\flushleft {\bf Acknowledgements}}

We acknowledge financial support from EPSRC Grant No. EP/P011713/1.

\bibliographystyle{elsarticle-num}
\bibliography{biblio}

\end{document}